\documentclass[prb,showpacs,preprintnumbers,amsmath,amssymb,twocolumn,superscriptaddress,longbibliography]{revtex4-1}
\usepackage{graphicx}
\usepackage{pbox}

\begin{document}
\title{Multigap superconductivity in the new BiCh$_{2}$-based layered superconductor La$_\mathrm{0.7}$Ce$_\mathrm{0.3}$OBiSSe} 

\author{A. Bhattacharyya}
\email{amitava.bhattacharyya@rkmvu.ac.in} 
\address{Department of Physics, Ramakrishna Mission Vivekananda Educational and Research Institute, Belur Math, Howrah 711202, West Bengal, India}
\author{D. T. Adroja} 
\affiliation{ISIS Facility, Rutherford Appleton Laboratory, Chilton, Didcot Oxon, OX11 0QX, United Kingdom} 
\affiliation{Highly Correlated Matter Research Group, Physics Department, University of Johannesburg, PO Box 524, Auckland Park 2006, South Africa}
\author{R. Sogabe}
\affiliation{Department of Physics, Tokyo Metropolitan University, 1-1 Minami-osawa, Hachioji, Tokyo 192-0397, Japan} 
\author{Y. Goto}
\affiliation{Department of Physics, Tokyo Metropolitan University, 1-1 Minami-osawa, Hachioji, Tokyo 192-0397, Japan} 
\author{Y. Mizuguchi}
\affiliation{Department of Physics, Tokyo Metropolitan University, 1-1 Minami-osawa, Hachioji, Tokyo 192-0397, Japan} 
\author{A. D. Hillier}
\affiliation{ISIS Facility, Rutherford Appleton Laboratory, Chilton, Didcot Oxon, OX11 0QX, United Kingdom} 

\date{\today} 

\begin{abstract}

\noindent The layered bismuth oxy-sulfide materials, which are structurally related to the Fe-pnictides/chalcogenides and cuprates superconductors, have brought substantial attention for understanding the physics of reduced dimensional superconductors. We have examined the pairing symmetry of recently discovered BiCh$_2$-based superconductor, La$_\mathrm{1-x}$Ce$_\mathrm{x}$OBiSSe with $x$ = 0.3,  through transverse field (TF) muon spin rotation measurement, in addition we present the results of magnetization,  resistivity and zero field (ZF) muon spin relaxation measurements. Bulk superconductivity has been observed below 2.7 K for $x$ = 0.3, verified by resistivity and magnetization data. The temperature dependence of the magnetic penetration depth has been determined from TF-$\mu$SR data can be described by an isotropic two-gap $s+s$ wave model compared to a single gap $s$- or anisotropic $s$-wave models, the resemblance with Fe-pnictides/chalcogenides and MgB$_2$. Furthermore, from the TF-$\mu$SR data, we have determined the London's penetration depth $\lambda_\mathrm{L}(0)$ = 452(3) nm, superconducting carrier's density $n_\mathrm{s}$ = 2.18(1) $\times$10$^{26}$ carriers/m$^{3}$ and effective mass enhancement $m^{*}$ = 1.66(1) $m_\mathrm{e}$, respectively. No signature of spontaneous internal field is found down to 100 mK in ZF-$\mu$SR measurement suggest that time-reversal symmetry is preserved in this system. 

\end{abstract}

\pacs{71.20.Be, 75.10.Lp, 75.40.Cx} 

\maketitle

\section{Introduction}

\noindent After the discovery of the BiCh$_{2}$-layered (Ch: S, Se) materials with the general formula REO$_\mathrm{1-x}$Fe$_\mathrm{x}$BiCh$_2$ (RE = rare earth) by Mizuguchi {\it et al. }~\cite{Mizuguchi1, Mizuguchi2, Mizuguchi3,Mizuguchirev1} and a superconducting transition temperature $T_\mathrm{C}$ up to 10 K, have drawn notable attention. This is mainly due to the strong spin-orbit coupling of Bi 6$p$ orbitals without local inversion symmetry with an anisotropic pairing symmetry~\cite{Mizuguchirev1}. The crystal structure consists of an alternate stacking of two BiCh$_{2}$ double layers and the LaO-blocking layers, which is similar to the Fe-As and Cu-O-Cu layers of the Fe-pnictides/chalcogenides and high-$T_\mathrm{C}$ cuprates~\cite{Bednorz, Kamihara} respectively. $T_\mathrm{C}$ can be increased by electron doping through replacing La by magnetic elements Ce, Pr, Yb, etc or through the external pressure, resembled with the Fe-based superconductors~\cite{Bhattacharyyarev}.  LaO$_\mathrm{1-x}$F$_\mathrm{x}$BiS$_\mathrm{2}$ is one of the most extensively studied system~\cite{Mizuguchirev1,Deguchi, J. Lee, Higashinaka1}, in which increasing F concentration, superconductivity is observed for $x \geq 0.2$, with a maximum $T_\mathrm{C}$ of 3.7 K for $x = 0.5$. This can be further enhanced, up to 10 K, by applying an external pressure of approximately 1 GPa~\cite{Mizuguchirev1,Mizuguchi4}. REO$_\mathrm{1-x}$F$_\mathrm{x}$BiS$_{2}$ (RE = La-Yb), Sr$_\mathrm{1-x}$La$_\mathrm{x}$FBiS$_{2}$, and La$_\mathrm{1-x}$Tr$_\mathrm{x}$OBiS$_{2}$ (Tr = Ti, Zr, Hf, and Th), were reported to be superconductors with $T_\mathrm{C}$ between 5 K and 10 K under applied pressure~\cite{Mizuguchirev1,Yazici,Jha1, Demura, Awana, Xing, Jha2}. \\ 

\noindent Hoshi {\it et al.}~\cite{Hoshi} reported the absence of isotope effects on $T_\mathrm{C}$ for LaO$_{0.6}$F$_{0.4}$Bi(S,Se)$_{2}$ sample with $^{76}$Se and $^{80}$Se isotopes indicating that phonons do not mediate the pairing. Recent angle-resolved photoemission spectroscopy~\cite{Ota} {of NdO$_{0.71}$ F$_{0.29}$BiS$_2$} indicate two-electron Fermi surfaces and the pairing symmetry is extremely anisotropic with nodelike feature. Field-angle-dependent Andreev reflection spectroscopy of LaO$_{0.5}$F$_{0.5}$BiS$_{2}$ suggest $d$-wave symmetry~\cite{Mizuguchirev1,Yazici}. On the other hand, the angular dependence of the upper critical field and $\mu$SR data suggest highly anisotropic multigap $s+s$-wave symmetry in LaO$_{0.5}$F$_{0.5}$BiS$_{2}$~\cite{Mizuguchirev1,Yazici}. Electrical resistivity measurements on CeO$_{0.5}$F$_{0.5}$BiS$_{2}$ under applied pressure display behavior consistent with a two-gap model~\cite{Mizuguchirev1,Sugimoto}. Neutron scattering investigations suggest that the electron-phonon coupling is much weaker and inferred the influence of charge fluctuations to mediate superconductivity in BiS$_2$-based superconductors~\cite{LeeINS,Yazici}. The non-linear Hall effect and magnetoresistance (which may be related to spin density wave  or charge density wave formation) as well as $\mu$SR and tunnel diode oscillator measurements on Bi$_{4}$O$_{4}$S$_{3}$ suggest multigap character and for Sr$_{0.5}$La$_{0.5}$FBiS$_{2}$ hint fully gapped $s$-wave state~\cite{Mizuguchirev1,Yazici, Lamura, BhattacharyyaBiS2}. In the case of CeOBiS$_2$, a logarithmic divergence of heat capacity, $C_\mathrm{P}\sim{-\ln(T)}$, suggest the system is very close to a quantum critical point~\cite{Higashinaka}. 

\noindent Density functional calculations on LaO$_{0.5}$F$_{0.5}$BiS$_{2}$ hint two-band electronic model with a strong nesting at ($\pi$, $\pi$, 0) with global $d_\mathrm{x^2-y^2}$ ($B_\mathrm{1g}$)-wave and anisotropic $s$-wave ($A_\mathrm{1g}$) symmetries~\cite{Mizuguchirev1,Yazici,Wan, Yildirim, Usui,Yang}. Renormalization-group calculations suggest that pairing symmetry in the BiCh$_2$-based superconductors is an admixture of singlets and triplets~\cite{Yazici}. Furthermore, the random phase approximation theory hints a spin and charge-fluctuation-mediated gap symmetry with extending $s$- or $d$-wave gap~\cite{Yazici}. The presence of Van Hove singularities and the logarithmic divergent of the density of states points towards unconventional pairing mechanism~\cite{Yazici}. The topology of the Fermi surface with electron and hole pockets of BiCh$_2$-based materials is then similar to Fe-pnictides/chalcogenides, which is believed to be multigap symmetry~\cite{Bhattacharyyarev}. A precise microscopic investigation is crucial to understand the controversial pairing symmetry of BiCh$_{2}$ based compounds. $\mu$SR is an indispensable method for examining the gap symmetry, pairing mechanism, and time-reversal symmetry breaking, facilitating an understanding of the unconventional superconductivity of Fe-pnictides/chalcogenides and cuprates high-$T_\mathrm{C}$ superconductors, which remain a puzzle~\cite{Sonier}. Herein, we investigate the gap symmetry and time reversal symmetry (TRS) of electron-doped, {by the mixed-valence state of Ce ions}, La$_{0.7}$Ce$_{0.3}$OBiSSe compound using TF- and ZF-$\mu$SR, respectively. Dome-shaped phase diagram is observed with the highest $T_\mathrm{C}$ of 3.1 K for $x$ = 0.3  in La$_{1-x}$Ce$_{x}$OBiSSe~\cite{Sogabe} resemblance with Eu$_{0.5}$La$_{0.5}$FBiS$_{2-x}$Se$_{x}$ compound with highest $T_\mathrm{C}$ is 3.8 K for $x$ = 0.8~\cite{Jinno}. From TF-$\mu$SR data, we found a multigap $s+s$-wave pairing symmetry of La$_{0.7}$Ce$_{0.3}$OBiSSe. 

\begin{figure*}[t]
\includegraphics[width=\linewidth]{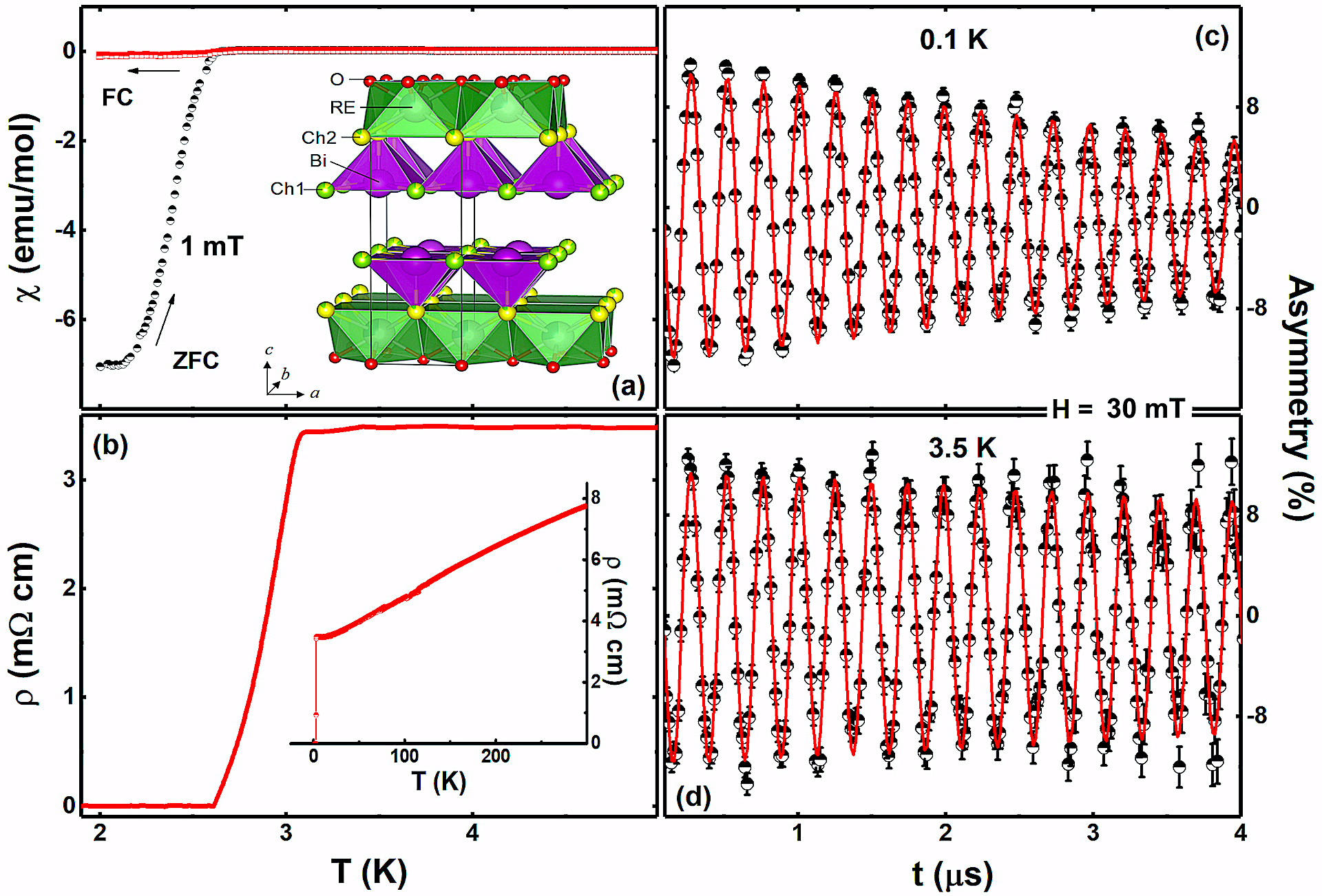}
\caption{ (a) represents the temperature dependence of the dc-susceptibility $\chi(T)$ of La$_{0.7}$Ce$_{0.3}$OBiSSe in an applied field 1 mT in zero field cool and field cool protocols. The inset exhibits a schematic illustration of the tetragonal crystal structure, where RE denotes the rare-earth site occupied by La and Ce. Ch1 and Ch2 denote the in-plane and out-of-plane chalcogen sites, respectively. The Ch1 site is largely occupied by Se, while the Ch2 site is occupied by S. (b) shows the temperature variation of the resistivity in the low-temperature limit in zero field. The inset presents resistivity data up to 300 K. Transverse-field muon asymmetry spectra as a function of time for La$_{0.7}$Ce$_{0.3}$OBiSSe collected at (c) $T$ = 0.1 K and (d) $T$ = 3.5 K in an applied field $H$ = 30 mT.}
\label{fig1}
\end{figure*}

\section{Experiment}

\noindent We have prepared a polycrystalline sample of La$_{0.7}$Ce$_{0.3}$OBiSSe via the solid-state method. Stoichiometric quantity of CeO$_{2}$(99.99\%), La$_{2}$O$_{3}$ (99.9\%), La$_{2}$S$_{3}$ (99.9\%), Ce$_{2}$S$_{3}$ (99.9\%), Bi (99.999\%), S (99.99\%) and Se (99.999\%) powders were put together and pressed into pellets, then sealed in an evacuated quartz tube, and heated at 973 K for 20 hours. The phase purity was determined using a powder X-ray diffraction Rigaku Miniflex diffractometer with Cu K$_\mathrm{\alpha}$ radiation. The temperature dependence of the magnetization measurements were carried out using a Superconducting Quantum Interference Device (SQUID magnetometer) with an applied field of 1 mT. The temperature dependence of resistivity was measured using a standard four-probe technique. 

\begin{figure}[t]
\includegraphics[width=\linewidth]{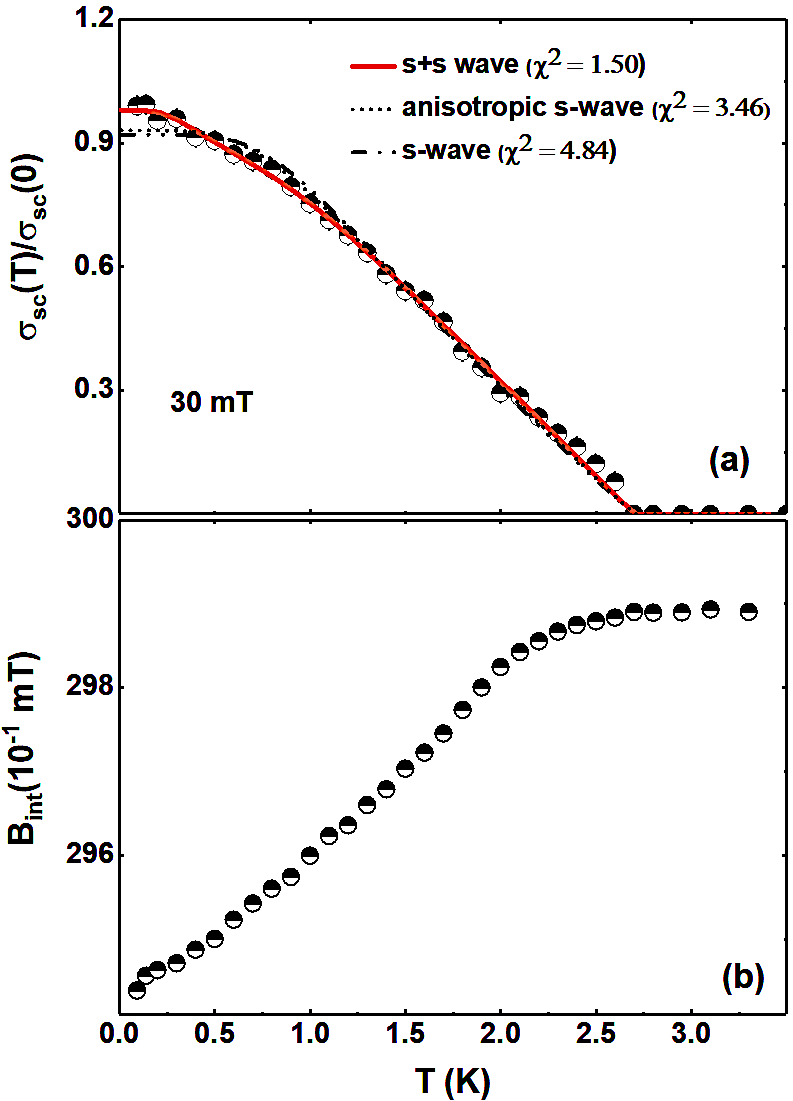}
\caption {(a) The temperature variation of the normalized superconducting depolarization rate $\sigma_\mathrm{sc}(T)/\sigma_\mathrm{sc}(0)$ with $H$ = 30 mT. The lines show the fits using $s$-wave (dashed dot blue), $s + s$-wave (solid red) and anisotropic $s$-wave (dot green) gap functions. Multigap behavior, with minimum $\chi^2$ = 1.5 for $s+s$-wave model, is confirmed  by the temperature dependence of magnetic penetration depth measurement. (b) The temperature variation of the internal field.}
\label{fig2}
\end{figure}  

\noindent $\mu$SR experiments in transverse field (TF)- and zero field (ZF)- modes, were carried out on the MUSR spectrometer at the ISIS Pulsed Neutron and Muon source, U.K. Powder sample of La$_{0.7}$Ce$_{0.3}$OBiSSe was mounted onto a silver (99.999\%) sample holder, and then it was inserted in a dilution refrigerator, which operated in the temperature range of 0.1 K $\leq$ T $\leq$ 4 K. Using active compensation coils, the stray magnetic fields at the sample position due to the Earth and neighboring instruments are canceled. TF-$\mu$SR was performed in the superconducting mixed state in an applied field of 30 mT, well above the lower critical field, $H_\mathrm{c1}$ = 1.6 mT. The asymmetry of the muon decay is calculated by $G(t) = [N_\mathrm{F}(t)- \alpha N_\mathrm{B}(t]/[N_\mathrm{F}(t)+ \alpha N_\mathrm{B}(t]$, where $\alpha$ is a constant determined from calibration measurements made in the paramagnetic state with a small (2 mT) applied transverse magnetic field. All the $\mu$SR data were analyzed using WiMDA software~\cite{Pratt}. 
  
\section{Results}

\subsection{Crystal Structure and Physical Properties}

\noindent La$_{0.7}$Ce$_{0.3}$OBiSSe crystallizes in a tetragonal structure with the space group of $P4/nmm$ (No: 129) as presented in the inset of Fig. \ref{fig1}(a). Se occupancy at the in-plane Ch1 site is higher than 85\%, intimating that Se-ions selectively occupy the in-plane Ch1 site, rather than out-of-plane Ch2 site is consistent with the observations in similar Eu$_{0.5}$La$_{0.5}$FBiS$_{2-x}$Se$_{x}$ and LaO$_{0.5}$F$_{0.5}$BiS$_{2-x}$Se$_{2}$ layered compounds~\cite{Hiroi, Goto}. The low-temperature magnetic susceptibility $\chi(T)$ data were measured in the Zero-field cooled (ZFC) and field-cooled (FC) protocol in an applied magnetic field of 1 mT is shown in Fig. \ref{fig1}(a), exhibit diamagnetic signals corresponding to the emergence of superconductivity at $T_\mathrm{C}$ = 2.7 K. Isothermal $M(H)$ data (not shown here) confirms bulk type-II superconductivity with a small lower critical field value of 1.6 mT. Fig. \ref{fig1}(b) demonstrates the temperature variation of the electrical resistivity $\rho(T)$, which manifests a sharp drop at 2.7 K consistent with the $\chi(T)$ data. The compound is metallic in its normal state, i.e., resistivity decreases with decreasing temperature $T$ down to 2.7 K, as shown in the inset of Fig. \ref{fig1}(b). {The metallic resistivity of La$_{0.7}$Ce$_{0.3}$OBiSSe originates from the mixed-valence of Ce ions ($\sim$ 3.47)~\cite{Sogabe}, which is in contrast to semiconducting behaviour observed in  Eu$_{0.5}$La$_{0.5}$FBiS$_{2-x}$Se$_{x}$ and LaO$_{0.5}$F$_{0.5}$BiS$_{2-x}$Se$_{2}$ compounds~\cite{Jinno}.} 

\subsection{TF-$\mu$SR analysis}

\noindent To examine the characteristics of the superconducting gap structure, we have carried out TF-$\mu$SR experiment. The TF-$\mu$SR data were collected in the field cooled state cooled in an applied field 30 mT. Typical spectra of the time dependence TF-$\mu$SR asymmetry both below and above $T_\mathrm{C}$ are shown in Fig.\ref{fig1}(c) and (d). Below $T_\mathrm{C}$, the asymmetry spectra decays due to the inhomogeneous field distribution of the flux line lattice. The TF-$\mu$SR spectra were best expressed utilizing damped Gaussian oscillatory decaying function,

\begin{eqnarray}
G_\mathrm{TF}(t) = A_\mathrm{s} \cos (2\pi\nu_\mathrm{1}t + \varphi)\exp (-\frac{\sigma^{2}t^{2}}{2})\exp(-\lambda t) \nonumber\\ +
 A_\mathrm{bg} \cos(2\pi\nu_\mathrm{2} t+\varphi)
\end{eqnarray} 

\noindent  where $\nu_\mathrm{1}$ and $\nu_\mathrm{2}$ are the frequencies of the muon spin precession from the sample and background Ag-sample holder, respectively. A$ _\mathrm{s}$ ($\sim$ 70~\%) and A$_\mathrm{bg}$ ($\sim$ 30~\%) are the initial asymmetries of the sample and background, $\varphi$ is the initial phase offset, $\sigma$ is the total Gaussian muon depolarization rate and $\lambda$ is the muon spin relaxation rate, which was added to account the electronic contribution from the Ce ions. The relaxation rate $\lambda$ is independent of temperature as revealed by ZF-$\mu$SR spectra, so we have fixed $\lambda$ ($\sim 0.014~\mu s^{-1}$) at its higher temperature value. In Eq. (1) the first term contains total sample relaxation rate $\sigma$, there are contributions from both the vortex lattice ($\sigma_\mathrm{sc}$) and nuclear dipole moments ($\sigma_\mathrm{nm}$, which is assumed to constant ($\sim$ 0.019 $\mu s^{-1}$) over the entire temperature range. The superconducting contribution to the muon relaxation rate is calculated using [$\sigma_\mathrm{sc} = \sqrt{\sigma^{2}-\sigma_\mathrm{nm}^{2}}$]. As $\sigma_\mathrm{sc}$ is directly, in the high $H_\mathrm{c2}$ limit, related to the superfluid density, we can model the temperature dependence of superfluid density using the following equation~\cite{AdrojaThFeAsN, BhattacharyyaThCoC2, AnandLaIrSi3, Prozorov, AdrojaK2Cr3As3, AdrojaCs2Cr3As3}

\begin{eqnarray}
\frac{\sigma_{sc}(T)}{\sigma_{sc}(0)} &=& \frac{\lambda^{-2}(T)}{\lambda^{-2}(0)}\\
 &=& 1 + \frac{1}{\pi}\int_{0}^{2\pi}\int_{\Delta(T)}^{\infty}(\frac{\delta f}{\delta E}) \times \frac{EdEd\phi}{\sqrt{E^{2}-\Delta(T})^2} \nonumber
\end{eqnarray}

\noindent where $f = [1+\exp(-E/k_\mathrm{B}T)]^{-1}$ is the Fermi function, $\phi$ is the azimuthal angle along the Fermi surface. The temperature and azimuthal angle dependent superconducting order parameter is $\Delta(T,\phi) = \Delta_\mathrm{0}\delta (T/T_\mathrm{C})g(\phi)$, where $\Delta_\mathrm{0}$ is the maximum gap value. The temperature dependence of the superconducting gap can be approximated by the relation $\delta(T/T_\mathrm{C}) = \tan\{1.82[1.018(T_\mathrm{C}/T-1)]^{0.51}\}$, g($\phi$) is the angular dependence of the superconducting gap structure, which is substituted by (a) 1-for isotropic $s$-wave gap [also for isotropic $s+s$ wave gap], (b) $\vert1+\cos(2\phi)\vert$/2 for an anisotropic $s$-wave~\cite{Pang2015, Annet1990}.\\
 
\noindent The temperature dependence of $\sigma_\mathrm{sc}(T)/\sigma_\mathrm{sc}(0)$, which is related to energy gap for quasi-particle excitations, was fitted with a single $s$-wave, anisotropic $s$-wave and $s+s$-wave models, and shown in Fig. \ref{fig2}(a). The temperature variation of the internal field determined from the fitting parameters of Eq. (1) is shown in Fig. \ref{fig2}(b), which decreases with decreasing temperature and is flat above $T_\mathrm{C}$ indicative of a superconducting transition. The $\sigma_\mathrm{sc}(T)$ increases with decreasing temperature confirms the presence of flux line lattice and suggest that London penetration depth decreases with decreasing temperature as $\sigma_\mathrm{sc} \propto 1/\lambda^{2}$. From the fit to $\sigma_\mathrm{sc}$ data, it is clear that the superconducting gap structure is best modeled by an isotropic $s+s$-wave model compared to a single $s$-wave model or an anisotropic $s$-wave model, which is agreement with the theoretical predictions of BiCh$_2$-based superconductors~\cite{Yazici}. The goodness to the fit $\chi^{2}$ = 1.5 value is lowest for $s+s$ model. The estimated parameters for the $s+s$ wave model show one larger gap $\Delta_\mathrm{1}(0)$ = 0.35(2) meV and a small gap $\Delta_\mathrm{2}(0)$ = 0.10(1) meV, which yield 2$\Delta(0)_\mathrm{1}$/k$_\mathrm{B}T_\mathrm{C}$ = 3.03. The fitting parameters obtained from isotropic $s$-wave, $s+s$-wave, and anisotropic $s$-wave models are summarized in Table~\ref{Tabel}. Multigap $s+s$-wave symmetry is also observed for Bi$_{4}$O$_{4}$S$_{3}$, LaO$_{0.5}$F$_{0.5}$BiS$_{2}$ and CeO$_{0.5}$F$_{0.5}$BiS$_{2}$ compounds~\cite{Yazici}. Other prominent examples of multigap superconductivity are represented by Fe-pnictides/chalcogenides~\cite{Bhattacharyyarev} Ba$_\mathrm{1-x}$K$_\mathrm{x}$Fe$_\mathrm{2}$As$_\mathrm{2}$, ThFeAsN, conventional BCS type MgB$_\mathrm{2}$~\cite{Rowell} and recently discovered on filled-skutterudite LaRu$_4$As$_{12}$~\cite{Juraszek}. Furthermore, similar small values for gap have been observed in Sr$_{0.5}$La$_{0.5}$FBiS$_{2}$, LaO$_{01-x}$F$_{x}$BiS$_{2}$ and La$_{1-x}$Y$_{x}$O$_{0.5}$F$_{0.5}$BiS$_{2}$ systems~\cite{Yazici}.

\begin{table}
\caption{Fitted parameters obtained from the fit to the
$\sigma_\mathrm{sc}$(T) data of La$_{0.7}$Ce$_{0.3}$OBiSSe using different gap models.}
\begin  {tabular}{|c|c|c|c|c| }
\hline
\pbox{4cm}{Model} &  g($\phi$)& \pbox{4cm} {Gap Value\\ $\Delta(0)$(meV)} & \pbox{4 cm}{Gap Ratio\\ 2$\Delta(0)$/k$_\mathrm{B}T_\mathrm{C}$} & $\chi^{2}$ \\ \hline
$s+s$-wave        & 1      & 0.35(2), 0.10(1)        & 3.03, 0.84 & 1.5 \\
anisotropic $s$-wave &$\frac{\vert 1+\cos 2\phi\vert}{2}$&0.38(1)&3.27 &3.5\\       
$s$-wave & 1 & 0.31(3) & 2.67& 4.8\\
\hline
\end{tabular}
\label{Tabel}
\end{table}  

\begin{figure}[t]
\includegraphics[width=\linewidth]{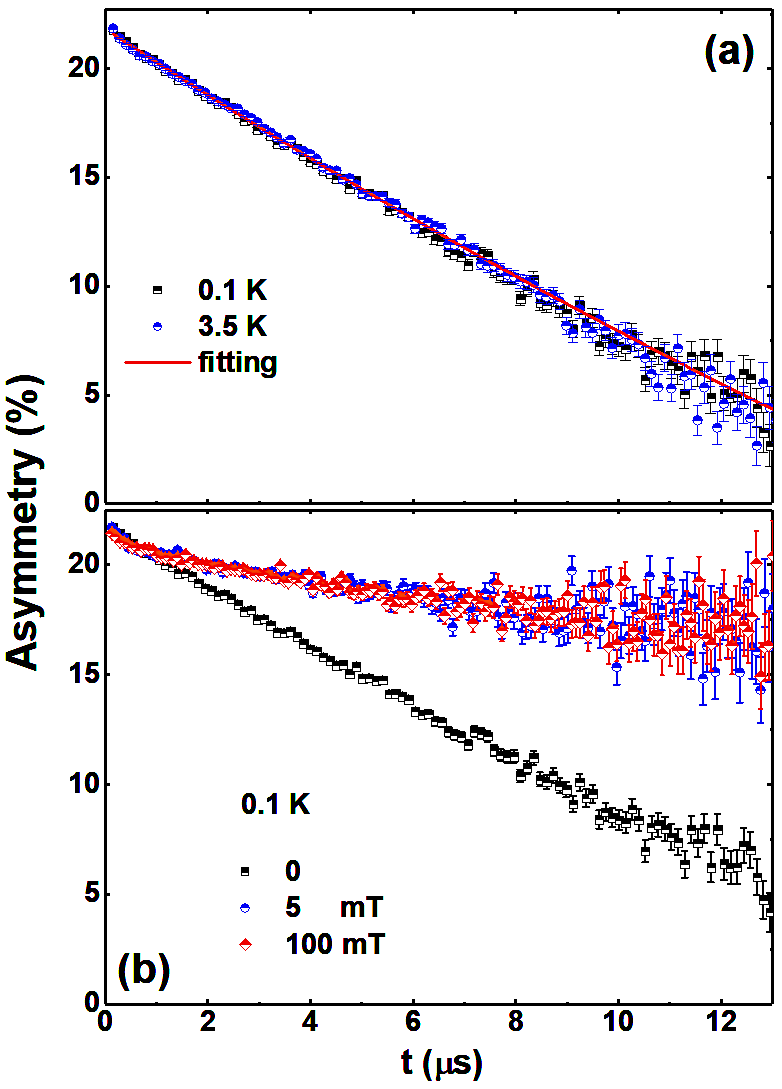}
\caption {(a) ZF-$\mu$SR time spectra of La$_{0.7}$Ce$_{0.3}$OBiSSe collected at 0.1 K (square), and 3.5 K (circle) are shown together with lines that are least-squares fits the data using Eq.~\ref{ZF}. (b) The LF-$\mu$SR time spectrum taken in an applied field of 5 mT and 100 mT at 0.1 K.}
\label{fig3}
\end{figure} 

\subsection{Superconducting Parameters}

\noindent Now we will discuss superconducting parameters of La$_{0.7}$Ce$_{0.3}$OBiSSe. For a triangular lattice, \cite{Sonier, Chia, Amato}

\begin{equation}
\frac{\sigma_\mathrm{sc}(T)^2}{\gamma_\mathrm{\mu}^2} = \frac{0.00371 \phi_\mathrm{0}^2}{\lambda^{4}(T)}
\end{equation}

\noindent here $\gamma_\mathrm{\mu}/2\pi$ is the gyromagnetic ratio (135.5 MHz/T) and $\phi_\mathrm{0}$ is the magnetic flux quantum (2.07$\times$10$^{-15}$T.m$^{2}$). Using London's theory~\cite{Sonier}

\begin{equation}
\lambda_\mathrm{L}^2 = \frac{m^{*}c^{2}}{4 \pi n_\mathrm{s} e^{2}}
\end{equation} 

\noindent where $m^{*} = (1+\lambda_\mathrm{e-ph})m_\mathrm{e}$ is the effective mass and $n_\mathrm{s}$ is the carrier density. $\lambda_\mathrm{e-ph}$ is the electron-phonon coupling strength, which can be estimated from Debye temperature ($\Theta_\mathrm{D}$) and $T_\mathrm{C}$ using McMillan's relation~\cite{McMillan,BhattacharyyaLaIr3,DasLaPt2Si2}

\begin{equation}
\lambda_\mathrm{e-ph} = \frac{1.04+\mu^{*}\ln(\Theta_\mathrm{D}/1.45T_\mathrm{C})}{(1-0.62\mu^{*})\ln(\Theta_\mathrm{D}/1.45T_\mathrm{C})-1.04}
\end{equation}

\noindent here $\mu^{*}$ is the repulsive screened Coulomb parameter usually assigned as $\mu^{*}$ = 0.13. For La$_{0.7}$Ce$_{0.3}$OBiSSe, we have $T_\mathrm{C}$ = 2.7 K and $\Theta_\mathrm{D}$ = 197 K, which gives $\lambda_\mathrm{e-ph}$ = 0.66. As La$_{0.7}$Ce$_{0.3}$OBiSSe is a type II superconductor we can assume that all normal state carriers ($n_\mathrm{e}$) contribute to superconductivity. We have estimated the magnetic penetration depth $\lambda_\mathrm{L}$(0) = 452(3) nm, superconducting carrier density $n_\mathrm{s}$ = 2.18(1) $\times$ 10$^{26}$ carriers/m$^{3}$, and effective-mass enhancement $m^{*}$ = 1.66(1) $m_\mathrm{e}$, respectively. The low value of carrier density is also observed for Bi$_4$O$_4$S$_3$~\cite{Yazici}.

\subsection{ZF-$\mu$SR analysis} 

\noindent To check the presence of any hidden magnetic ordering or broken time-reversal symmetry in La$_{0.7}$Ce$_{0.3}$OBiSSe, we have carried out ZF-$\mu$SR experiment. The time evolution of ZF-$\mu$SR spectra below and above $T_\mathrm{C}$ is shown in Fig. \ref{fig3}(a) for $T$ = 0.1 K and $T$ = 3.5 K. The absence of muon precession or loss of initial asymmetry value at $t$ = 0, excludes the presence of a large internal magnetic field, as seen in a magnetically ordered compound or magnetic impurity. Moreover, the only possibility is that the muon-spin relaxation is due to static, randomly oriented local fields associated with the nuclear moments at the muon site and a weak contribution from the Ce electronic moments. The ZF-$\mu$SR data can be well described using a damped Gaussian Kubo-Toyabe function,

\begin{equation}
G_\mathrm{ZF}(t) = A_\mathrm{3}G_\mathrm{KT}(t)\exp^{-\lambda_\mathrm{\mu}t}+A_\mathrm{ZF}
\label{ZF}
\end{equation}  

here 

\begin{equation}
G_\mathrm{KT}(t) = [\frac{1}{3}+\frac{2}{3}(1-\sigma_\mathrm{KT}^{2}t^{2})e^{\frac{-\sigma_\mathrm{KT}^2t^2}{2}}]
\end{equation}

\noindent  the Gaussian Kubo-Toyabe function, $A_\mathrm{3}$ is the initial asymmetry, $\lambda_\mathrm{\mu}$ is the relaxation rate, and $A_\mathrm{bg}\sim$ 29\% is the background signal. $A_\mathrm{3}$, $A_\mathrm{ZF}$ and $\sigma_\mathrm{KT}$ are found to be nearly independent of temperature. No significant change is observed in the relaxation rate at 0.1 K ( below $T_\mathrm{C}$) and 3.5 K (above $T_\mathrm{C}$ ), reveals that time reversal symmetry is preserved in the superconducting state. By fitting the time evolution of ZF-$\mu$SR spectra with Equation~\ref{ZF} we get $\sigma_\mathrm{KT}$ = 0.06(1) $\mu$s$^{-1}$ and $\lambda_\mathrm{\mu}$ = 0.046(2) $\mu$s$^{-1}$ at $T$ = 0.1 K and $\sigma_\mathrm{KT}$ = 0.06(1) $\mu$s$^{-1}$ and $\lambda_\mathrm{\mu}$ = 0.043(3) $\mu$s$^{-1}$ at $T$ = 3.5 K. The red solid line shows the fitting to the experimental data. Since within error bars the difference of $\lambda_\mathrm{\mu}$ and $\sigma_\mathrm{KT}$ at $T \geqslant$ $T_\mathrm{C}$ and $T \leqslant T_\mathrm{C}$ are negligible, implies no time reversal symmetry breaking in La$_{0.7}$Ce${0.3}$OBiSSe and interesting no change to the Ce magnetic moment dynamics. A small longitudinal field removes any relaxation from spontaneous fields {see Fig.~\ref{fig3}(b). For La$_{0.7}$Ce$_{0.3}$OBiSSe, 5 mT is sufficient to decoupled to nuclear moments from the relaxation channel.

\section{Conclusion}

\noindent In conclusion, we have performed the resistivity, magnetization, ZF- and TF-$\mu$SR measurements to investigate the superconductivity of the BiCh$_{2}$-based layered superconductor La$_{0.7}$Ce$_{0.3}$OBiSSe. Resistivity and magnetization data confirm the bulk nature of superconductivity at 2.7 K. Temperature dependence of magnetic penetration depth best modeled by isotropic $s+s$ wave compared to single gap isotropic $s$-wave or anisotropic $s$-wave models, which agrees with multigap $s+s$-wave gap for Bi$_{4}$O$_{4}$S$_{3}$, LaO$_{0.5}$F$_{0.5}$BiS$_{2}$, and CeO$_{0.5}$F$_{0.5}$BiS$_{2}$ compounds and two-band electronic model suggested by theoretical calculations~\cite{Yazici}. The observed gap symmetry is a resemblance to Fe-pnictides/chalcogenides and MgB$_2$.  ZF-$\mu$SR measurement designates no spontaneous magnetic field below $T_\mathrm{C}$. The absence of a spontaneous magnetic field below $T_\mathrm{C}$ indicates the TRS is preserved. Our TF-$\mu$SR result will help to understand the contradictory results on the superconducting pairing mechanisms for the BiCh$_{2}$-based layered superconductor. 

\subsection*{Acknowledgment}

\noindent AB gratefully acknowledges the financial support from the Department of Science and Technology, India (SR/NM/Z-07/2015) for the financial support and Jawaharlal Nehru Centre for Advanced Scientific Research (JNCASR) for managing the project and the Department of Science and Technology (DST) India, for an Inspire Faculty Research Grant (DST/INSPIRE/04/2015/000169). DTA would like to thank the Royal Society of London for the UK-China Newton funding and the Japan Society for the Promotion of Science for an invitation fellowship and STFC/UKRI for beamtime.

\end{document}